\documentclass[aps,prl,twocolumn,10pt,superscriptaddress]{revtex4-1}
\usepackage{amssymb,amsthm,amsmath,amsfonts}
\usepackage{xcolor,graphicx,ulem}
\usepackage[colorlinks=true,urlcolor=blue,citecolor=blue,linkcolor=blue]{hyperref}

\begin{document}

\title{Gaussian versus non-Gaussian filtering of phase-insensitive nonclassicality}

\author{B. K\"uhn}
\author{W. Vogel}
\affiliation{Arbeitsgruppe Quantenoptik, Institut f\"ur Physik, Universit\"at Rostock, D-18051 Rostock, Germany}

\author{V. Thiel}\email{oqt@uoregon.edu}
\affiliation{Clarendon Laboratory, University of Oxford, Parks Road, Oxford, OX1 3PU, UK}
\affiliation{Department of Physics and Oregon Center for Optical, Molecular, and Quantum Science, University of Oregon, Eugene, Oregon 97403, USA}

\author{S. Merkouche}
\affiliation{Department of Physics and Oregon Center for Optical, Molecular, and Quantum Science, University of Oregon, Eugene, Oregon 97403, USA}

\author{B. J. Smith}
\affiliation{Department of Physics and Oregon Center for Optical, Molecular, and Quantum Science, University of Oregon, Eugene, Oregon 97403, USA}
\date{\today}

\begin{abstract}
	Measures of quantum properties are essential to understanding the fundamental differences between quantum and classical systems as well as quantifying resources for quantum technologies. 
	Here two broad classes of bosonic phase-space functions, which are filtered versions of the Glauber-Sudarshan $P$~function, are compared with regard to their ability to uncover nonclassical effects of light through their negativities.
	Gaussian filtering of the $P$~function yields the family of $s$-parametrized quasiprobabilities, while more powerful regularized nonclassicality quasiprobabilities are obtained by non-Gaussian filtering.
	A method is proposed to directly sample such phase-space functions for the restricted case of phase-independent quantum states from balanced homodyne measurements. 
	This overcomes difficulties of previous approaches that manually append uniformly distributed optical phases to the measured quadrature data.	
	We experimentally demonstrate this technique for heralded single- and two-photon states using balanced homodyne detection with varying efficiency.
	The $s$-parametrized quasiprobabilities, which can be directly sampled, are non-negative for detection efficiencies below 0.5.
	By contrast, we show that significant negativities of non-Gaussian filtered quasiprobabilities uncover nonclassical effects for arbitrarily low efficiencies.	
\end{abstract}
\maketitle

\paragraph*{Introduction.---}\hspace{-3ex}
	The distinction between quantum and classical properties of a physical system has played a fundamental role in the development of quantum theory since its inception \cite{wheeler83}. 
	Tools to probe the boundary between the classical and quantum domains have developed over more than a century since the foundations of quantum theory were set out. 
	These techniques have become increasingly important in quantum information science, where the ability to quantify nonclassical, i.e. quantum, properties of a physical system determines how well the system can perform a particular technological task \cite{Chitambar2019}. 
	Phase-space distributions have emerged as canonical representations of quantum systems that can be utilized to distinguish their nonclassical properties \cite{Munroe95, Sperling2020}. 
	
	There are a number of distinguishing characteristics displayed by quantum states of light. For single-mode fields, such non-classical traits include non-Gaussian Wigner representations and singular Glauber-Sudarshan P-function. These reflect different characteristics of quantum states of light, which have different utility in quantum information science and technology.	In 1963, it was discovered that all states of a single electromagnetic field mode can be represented in the form~\cite{Glauber1963,Sudarshan1963}
	\begin{align}\label{eq:GS}
		\hat\rho=\int d^2\alpha\,P(\alpha)|\alpha\rangle\langle\alpha|,
	\end{align}
	by means of the Glauber-Sudarshan $P$~function $P(\alpha)$, which contains complete information about a state $\hat{\rho}$.
	Coherent states $|\alpha\rangle$ are known to be the only pure states with a non-negative $P$~function \cite{Hillery1995} and can be considered as the analogue of a classical radiation field of amplitudes $\alpha$.
	The $P$~function of a statistical mixture of coherent states is thus the same as the classical phase-space distribution of the corresponding statistical distribution over classical amplitudes.
	Accordingly, it is reasonable to define a state as classical if it has a representation as in Eq.~\eqref{eq:GS} with a classical probability distribution $P(\alpha)=P_{\mathrm{cl}}(\alpha)$~\cite{Titulaer1965}. 
	A remarkable aspect of quantum physics is that it allows states that cannot be represented by a completely positive $P$~function and thus are called nonclassical states of the field~\cite{Mandel1986}. 
	Prominent examples of nonclassical states with negative $P$~function values include single-photon states and squeezed states.
	It would be an easy task to experimentally certify nonclassicality if the $P$~function of all states existed as a regular function, but the opposite is the case.
	In fact, this function can even contain infinite derivatives of the Dirac $\delta$ distribution~\cite{Sperling2016}.
	Accordingly, the $P$~function is in general not accessible experimentally.
	An established approach to uncover nonclassicality in phase space is by convolving the $P$~function with a Gaussian function to transform it into a regular function, so-called s-parametrized quasiprobabilities, whose negativities unambiguously represent nonclassicality of the state~\cite{Cahill1969}.
	A drawback of this method is that many states, such as the important class of squeezed Gaussian states, are not identified as nonclassical, since their Gaussian-regularized $P$~function is nonnegative. 
	In addition, such approaches to identifying nonclassicality of states place strict requirements on measurement efficiency.
	For this reason, non-Gaussian filter functions with specific properties have been introduced~\cite{Kiesel2010}.
	Filtering the $P$~function with non-Gaussian functions allows a complete nonclassicality test, as the strength of filtering can be arbitrarily reduced while preserving the regularity of the resulting nonclassicality quasiprobability distribution, and benefits from reduced sensitivity to detector efficiency.

	In the present Letter, we study the benefits of non-Gaussian compared to Gaussian filtered $P$~functions to uncover nonclassical effects in the realistic scenario of low quantum efficiency detection on the basis of experimental quadrature data of lossy single- and two-photon states. 
	Experimental access to regular phase-space representations is provided by balanced homodyne detection (BHD) of light~\cite{Yuen1983,Smithey1993,Welsch1999}, which measures the quadrature statistics of the electric field strength of the radiation field.
	In Refs.~\cite{Kiesel2011_2,Agudelo2015} direct sampling formulas were introduced to easily obtain quasiprobabilities from phase-sensitive BHD data. 
	We go beyond this by developing direct sampling formulas of non-Gaussian regularized $P$~functions from phase-insensitive quadrature measurements via BHD.
	This method is then applied to BHD of heralded single- and two-photon states for different detection efficiencies.

\paragraph*{Regular phase-space functions.---}\hspace{-3ex}
	Singularities of the Glauber-Sudarshan $P$~function dictate alternative strategies to experimentally access nonclassicalities of quantum states.
	The $P$~function is the Fourier transform 
	\begin{align}
		P(\alpha)=\dfrac1{\pi^2}\int d^2\beta\,e^{\alpha\beta^\ast-\alpha^\ast\beta}\Phi(\beta),
	\end{align}
	of the characteristic function $\Phi(\beta)$.
	The latter can grow maximally as $e^{|\beta|^2/2}$; see Ref.~\cite{Sperling2016} for details.
	Accordingly its Fourier transform $P(\alpha)$ does in this case not exist as a regular function.	
	
	To obtain regular phase-space functions the characteristic function can be multiplied by a filter function $\Omega(\beta)$ which decays stronger than the Gaussian $e^{-|\beta|^2/2}$, resulting in a new phase-space function
	\begin{align}\label{eq:POmega}
		P_\Omega(\alpha)=\dfrac1{\pi^2}\int d^2\beta\,e^{\alpha\beta^\ast-\alpha^\ast\beta}\Omega(\beta)\Phi(\beta).
	\end{align}
	If the Fourier transform of the filter function is non-negative, the regularization procedure does not introduce negativities in $P_\Omega(\alpha)$.
	Accordingly, a negativity of the latter unambiguously certifies nonclassicality of the state.
	Non-Gaussian filter functions which fulfill these requirements were introduced in Ref.~\cite{Kiesel2010} as autocorrelations
	\begin{align}\label{eq:Omegaq}
		\Omega^{(q)}_w(\beta)=\int d^2\gamma\,\chi^{(q)\ast}_w(\gamma)\,\chi^{(q)}_w(\beta+\gamma)
	\end{align}
	of functions of the form
	\begin{align}\label{eq:smallomega}
		\chi^{(q)}_w(\beta)=\dfrac1{w}2^{1/q}\sqrt{\dfrac{q}{2\pi\Gamma(2/q)}}\exp\left[-\left(\dfrac{|\beta|}{w}\right)^q\right],
	\end{align}
	where $2<q<\infty$, $w$ is a positive value, and $\Gamma(\cdot)$ is the gamma function.
	The parameter $w$ controls the width of the filter and thus the degree of filtering, which affects how smooth the resulting quasi-probability distribution becomes.
	For $w\to\infty$ one obtains the original $P$~function.
 	In the limit $q\to\infty$ the filter function in Eq.~\eqref{eq:Omegaq} has the analytical form~\cite{Kuehn2014}
 	\begin{equation}\label{eq:FilterInfty}
		\Omega(\beta)=\dfrac{2}{\pi}\left[\arccos\left(\dfrac{|\beta|}{2w}\right)-\dfrac{|\beta|}{2w}\sqrt{1-\dfrac{|\beta|^2}{4w^2}}\right]\mathrm{rect}\left(\dfrac{|\beta|}{4w}\right),
	\end{equation}
	where $\mathrm{rect}(x)$ is one for $x\leq 1/2$ and zero otherwise.
	In the opposite limiting case $q=2$, the function in Eq.~\eqref{eq:Omegaq} is essentially the autocorrelation of two Gaussians and, therefore, it reduces to a Gaussian function.
	Rescaling the parameter $w$ according to $s=1-1/w^2$ with $s\leq 1$, yields the filter function
	\begin{align}\label{eq:sfilter}
		\Omega(\beta)&=\exp\left(-\dfrac{1-s}{2}|\beta|^2\right).
	\end{align}
	Inserting this Gaussian filter in Eq.~\eqref{eq:POmega}, the established class of $s$-parametrized quasiprobabilities $P(\alpha;s)$ is retrieved~\cite{Cahill1969}. 
	To guarantee regularity of these quasiprobabilities $s$ must be chosen less-equal to $0$.
	In other words, only $s$-parametrized quasiprobabilities at least as smooth as the Wigner function, $W(\alpha)=P(\alpha;0)$, are regular for all quantum states~\cite{Wigner1932}.
	It is known that for $s\leq -1$ the corresponding quasiprobability is always non-negative~\cite{Husimi1940}, i.e., nonclassicality cannot be identified by negativities in this range.
	Consequently, this family of phase-space functions is useful only for $-1<s\le 0$ to identify nonclassicality through negative values. 
	Accordingly, only a subset of all states is  uncovered to be nonclassical by these quasiprobabilities.
	Note that the important class of the squeezed vacuum states is not included in this set.
	A further disadvantage of the $s$-parametrized quasiprobability is the inability to certify nonclassicality in the presence of high constant losses or low detection efficiency $\eta$.
	In particular, all states detected with $\eta\le 0.5$ have nonnegative $s$-parametrized quasiprobabilities for $s\le0$.
		
	By contrast, filters with $q>2$ in Eq.~\eqref{eq:Omegaq} provide a full nonclassicality test, since the corresponding filtered function $P_\Omega$ is always regular for arbitrarily large parameter $w$.
	We want to point out here that an arbitrarily small quantum efficiency $\eta$ can be compensated by a rescaled larger filter parameter $w/\sqrt{\eta}$.
	That is why these filters are referred to as nonclassicality filters and the functions $P_\Omega$ are denoted as nonclassicality quasiprobabilities~\cite{Kiesel2010}. 
	It is a central purpose of this work to demonstrate the power of these nonclassicality quasiprobabilities to certify nonclassical effects in the presence of high losses ($\eta<0.5$), where the $s$-parametrized quasiprobabilities fail to display any nonclassicality.

\paragraph*{Direct sampling of quasiprobabilities.---}\hspace{-3ex}	
	The phase-dependent statistics $p(x;\varphi)=\langle x;\varphi | {\hat{\rho}} | x;\varphi \rangle$ of the quadrature $x$, which is proportional to the electric field strength of the electromagnetic radiation at the optical phase $\varphi$, contains full information about a quantum state ${\hat{\rho}}$.
	Accordingly there exists a unique mapping from this probability distribution to the quasiprobabilities in Eq.~\eqref{eq:POmega}.
	In particular, both quantities are connected by the relation~\cite{Kiesel2011_2}
	\begin{align}\label{eq:Pfromp}
		P_\Omega(\alpha)=\int_{-\infty}^\infty dx\int_0^{2\pi} d\varphi\,\dfrac{p(x;\varphi)}{2\pi}\,f_\Omega(x,\varphi;\alpha),
	\end{align}
	via the pattern function
	\begin{align}\label{eq:pattern}
		f_\Omega(x,\varphi;\alpha)=\dfrac{2}{\pi}\int_0^\infty db\,b\,e^{b^2/2}\Omega(b)\cos[\xi(x,\varphi,\alpha)b],
	\end{align}
	where
	\begin{align}\label{eq:xi}
		\xi(x,\varphi,\alpha)=x+2|\alpha|\sin(\arg(\alpha)+\varphi-\pi/2).
	\end{align}	
	The quadrature distribution $p(x;\varphi)$ can be measured experimentally using BHD~\cite{Yuen1983,Smithey1993,Welsch1999}.
	BHD results in a set of $N$ statistically-independent quadrature-phase pairs $\{(x_j,\varphi_j)\}_{j=1,\dots,N}$, where the phases $\varphi_j$ must be scanned uniformly in the range $[0,2\pi)$, ensuring that the quadrature measurements are properly averaged over the optical phases. The values of $x_j$ depend upon the state and the number of measurement outcomes improves the estimate of nonclassicality by reducing errors from statistical fluctuations; see Ref.~\cite{Agudelo2015} for further details. Equation~\eqref{eq:Pfromp} allows one to formulate the convenient direct sampling formula
	\begin{align}\label{eq:phasesensitivesampling}
		P_\Omega(\alpha)\approx\dfrac1{N}\sum_{j=1}^Nf_\Omega(x_j,\varphi_j;\alpha),
	\end{align}
	to estimate the regular phase-space distribution.
	The statistical uncertainty of the estimate in Eq.~\eqref{eq:phasesensitivesampling} is straightforwardly calculated to be
	\begin{align}\label{eq:sigma}
		\sigma\left\{P_\Omega(\alpha)\right\}=\dfrac1{\sqrt{N(N-1)}}\sqrt{\sum_{j=1}^N\left[f_\Omega(x_j,\varphi_j;\alpha)-P_\Omega(\alpha)\right]^2}.
	\end{align}
	Negativities of $P_\Omega$ certify nonclassicality, therefore, we evaluate the signed statistical significance
	\begin{align}\label{eq:signif}
		\Sigma=\min_\alpha\left[\dfrac{P_{\Omega}(\alpha)}{\sigma\left\{P_\Omega(\alpha)\right\}}\right]
	\end{align}
	to ensure reliable results.
	The direct sampling of nonclassicality quasiprobabilities was successfully demonstrated for a squeezed vacuum state~\cite{Kiesel2011_2}.
		
	In some situations it is known {\it{a priori}} that the state of the field is phase invariant, i.e., 
	\begin{align}\label{eq:phaseinsensitivequadrature}
		p(x;\varphi)=p(x).
	\end{align}
	A prominent example are the Fock states.
	Generally, the set of states belonging to the category of phase-insensitive states is formed by statistical mixtures of Fock states.
	In this case it is not necessary to record the optical phase $\varphi_j$ associated with the measured quadrature $x_j$.
	Rather, only a set of $N$ quadratures $\{x_j\}_{j=1,\dots,N}$ is recorded.
	One can still use Eq.~\eqref{eq:phasesensitivesampling} if a phase $\varphi_j$ is randomly attributed---according to a uniform distribution---to each quadrature sample $x_j$.
	However, this can lead to ambiguity problems, particularly when $N$ is small.
	Thus, we derive a pattern function, for the case of phase-insensitive quantum states that only depends on the quadrature $x$.
	Inserting Eq.~\eqref{eq:phaseinsensitivequadrature} into Eq.~\eqref{eq:Pfromp}, one obtains
	\begin{align}\label{eq:Ppx}
		P_\Omega(\alpha)=\int_{-\infty}^\infty dx\,p(x)\,\overline{f}_\Omega(x;\alpha),
	\end{align}
	with a new pattern function which reads as
	\begin{align}\label{eq:phaseinsensitivepattern}
		\overline{f}_\Omega(x;\alpha)=\dfrac1{2\pi}\int_0^{2\pi}d\varphi\,f_\Omega(x,\varphi;\alpha).
	\end{align}
	On the basis of Eq.~\eqref{eq:Ppx} it is straightforward to show that the phase-space distribution can be estimated by a direct sampling formula of the form
	\begin{align}\label{eq:directinsensitive}
		P_\Omega(\alpha)\approx\dfrac1{N}\sum_{j=1}^N\overline{f}_\Omega(x_j;\alpha),
	\end{align}
	which no longer contains phase values compared to Eq.~\eqref{eq:phasesensitivesampling}. 
	The statistical error of the estimate in Eq.~\eqref{eq:directinsensitive} is given by
	\begin{align}\label{eq:phaseindependentsigma}
		\sigma\left\{P_\Omega(\alpha)\right\}=\dfrac1{\sqrt{N(N-1)}}\sqrt{\sum_{j=1}^N\left[\overline{f}_\Omega(x_j;\alpha)-P_\Omega(\alpha)\right]^2}.
	\end{align}
	Now, we further evaluate the integral in Eq.~\eqref{eq:phaseinsensitivepattern}.
	Using Eq.~\eqref{eq:pattern}, it holds
	\begin{align}
		\overline{f}_\Omega(x;\alpha)=\dfrac{2}{\pi}\int_0^\infty \!db\,b\,e^{b^2/2}\Omega(b)\dfrac1{2\pi}\int_0^{2\pi}\!\!d\varphi\,\cos[\xi(x,\varphi,\alpha)b].
	\end{align}
	Recalling the definition in Eq.~\eqref{eq:xi}, and performing the phase integration,  yields
	\begin{align}\label{eq:pat}
		\overline{f}_\Omega(x;\alpha)=\dfrac{2}{\pi}\int_0^\infty db\,b\,e^{b^2/2}\Omega(b)\,J_0(2|\alpha|b)\cos(xb)
	\end{align}
	with $J_0(\cdot)$ being the Bessel function of the first kind. 
	The expression for the phase-insensitive pattern function in Eq.~\eqref{eq:pat} together with the direct sampling formulas in Eqs.~\eqref{eq:directinsensitive} and~\eqref{eq:phaseindependentsigma} are central results of the present work.
		
	Note that the nonclassicality quasiprobability of a phase-independent state, namely a single-photon-added thermal state, has been reconstructed from experimental data~\cite{Kiesel2011_1}.
	However, this was performed in an indirect manner by first sampling the characteristic function of the $P$~function and on this basis the nonclassicality quasiprobability was determined. This required optimization of the filter function for the state that was studied. The resulting filter function did not have an analytical form and the error calculation in this scenario becomes cumbersome already for the single-mode case. In the present manuscript, the analytical representation is a significant improvement in the methodology to certify non-classicality with the best statistical significance for the considered states.
	
	As already stated in the preceding section, in the case of the Gaussian filter in Eq.~\eqref{eq:sfilter} the parameter $s$ must be smaller than zero in order to apply the direct sampling formula, since otherwise the pattern functions in Eqs.~\eqref{eq:pattern} and~\eqref{eq:pat} do not exist as regular expressions.
	Beneficially, such a confinement does not exist in the case of the nonclassicality filters in Eq.~\eqref{eq:Omegaq} for $q>2$.
	In Ref.~\cite{Kuehn2014} it was shown that among the possible values of the parameter $q$ the filter in Eq.~\eqref{eq:FilterInfty}, which corresponds to $q\to\infty$, requires the minimal amount $N$ of quadrature data points to significantly certify nonclassicality on the basis of the negativities of $P_\Omega$.
	For this reason, we will apply this particular nonclassicality filter in the following considerations. 

\begin{figure}[t]
		\centering
		\includegraphics[width=.99\linewidth]{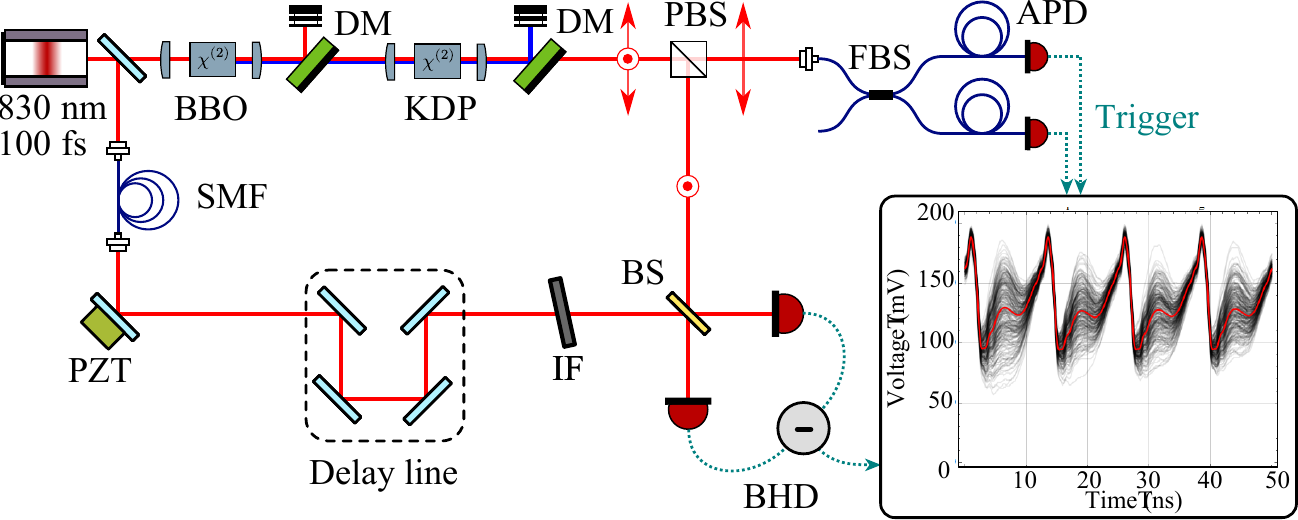}
		\caption{Experimental scheme. PBS: Polarizing beam-splitter. APD: Avalanche Photodiode. PZT: Piezo-electric actuator. IF: Interference filter. BHD: Balanced homodyne detection. BS: Beam-splitter. FBS: Fiber beam-splitter. Inset: 250 acquisitions of four pulses from the BHD with the residual mean voltage.}
		\label{fig:exp}
	\end{figure}

\paragraph*{Experimental setup.---}\hspace{-3ex}
	Single- and two-photon states were generated by degenerate, co-linear, type-II spontaneous parametric down conversion (SPDC) in potassium di-hydrogen phosphate (KDP) \cite{mosley2008, Cooper2013_2, Davis2018}. 
	The SPDC process was pumped by 3.5 nm bandwidth pulses centered at 415 nm wavelength produced by second-harmonic generation from a Ti:Sapphire laser in beta-barium borate (BBO), as depicted in Fig. \ref{fig:exp}. 
	The idler beam was coupled into a single-mode fiber and sent to spatially multiplexed single-photon counting modules (SPCMs).
	A single (double) click detection event in the multiplexed SPCMs heralds a single- (two-) photon state in the signal path.
	The signal beam was sent to a BHD for quadrature measurements.

	The BHD \cite{Cooper2013} utilized a local oscillator (LO) pulse train derived from the Ti:Sapphire laser system mode matched to the signal photons. 
	The maximum overall efficiency of the system was determined to be $\eta \approx 0.4$ (see supplementary information). 
	The efficiency of the detection was set to approximately $0.4$, $0.3$, $0.2$, $0.1$ and $0.05$, achieved by tuning the mode overlap between the signal photons and the LO, and data for heralded single- and two-photon states collected at each setting.
	We sampled $8 \cdot 10^5$ and $6 \cdot 10^5$ quadrature values for single- and two-photon states, respectively. 
	The heralding rate was 500 kHz in the former case and 500 Hz in the latter.
	Further details about the experimental setup can be found in the supplementary information ~\cite{supp}.
	
\begin{figure}[t]
	\centering
	\includegraphics[width=0.8\linewidth]{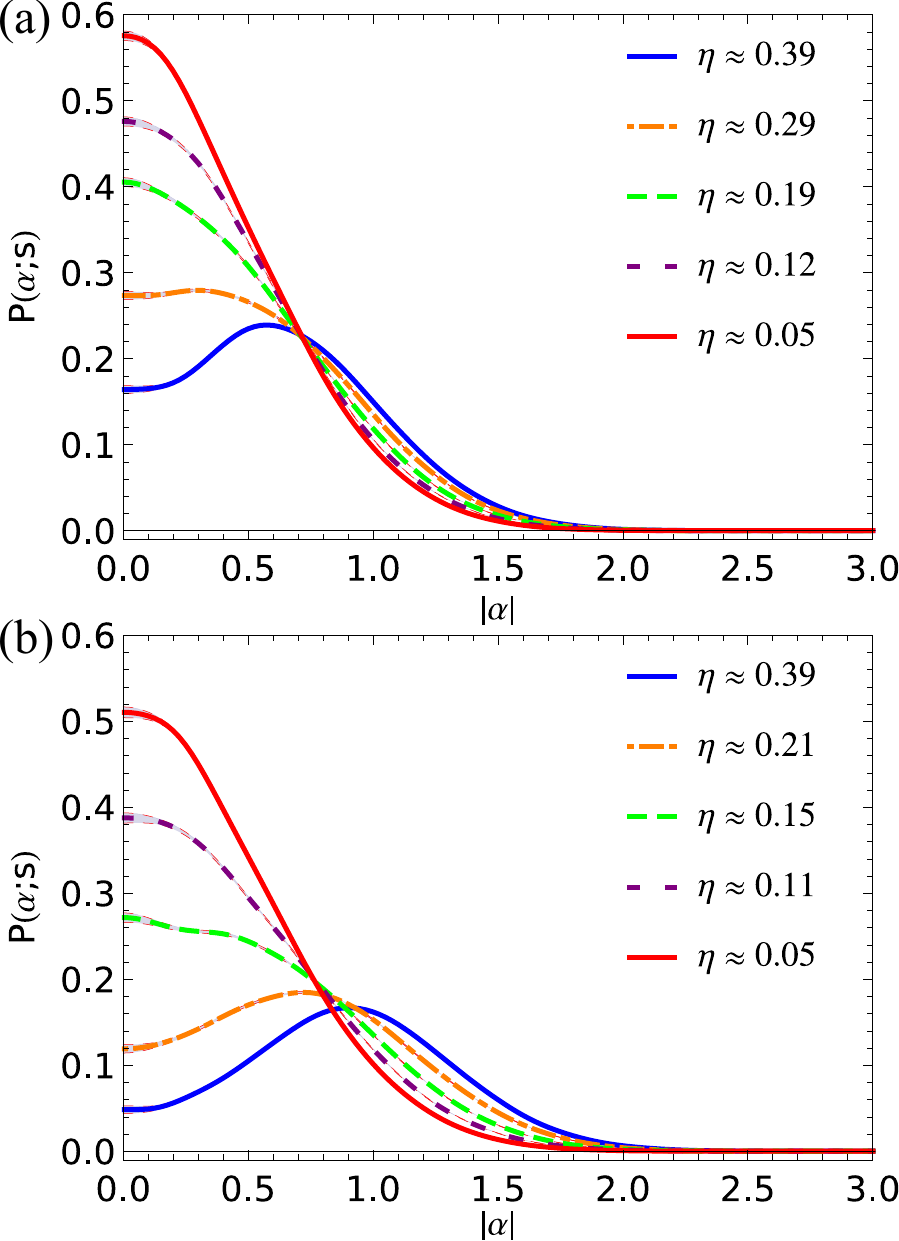}	
	\caption{
		(Color online)
		The directly sampled $s$-parametrized quasiprobabilities as a function of $|\alpha|$ for $s=-0.04$ and five different quantum efficiencies.
		(a) Single photon; (b) Two photons.
		The thin dashed line correspond to an error of one standard deviation, which is barely noticeable.
	}
	\label{fig:spara}
\end{figure}

\paragraph*{Comparing Gaussian with non-Gaussian filtering.---}\hspace{-3ex}	
	In this section, we apply the phase-independent direct sampling formula in Eq.~\eqref{eq:directinsensitive} together with Eq.~\eqref{eq:phaseindependentsigma} to the measured quadrature data sets. 
	First we sampled the $s$-parametrized quasiprobabilities, which is only possible for $s<0$.
	Figure~\ref{fig:spara} (a) and~(b) show these phase-space functions for the single- and two-photon state, respectively, for an $s$-parameter close to $0$.
	As expected for the low efficiencies $\eta<0.5$ under consideration, no negativities appear.
	This once more shows the inability of the $s$-parametrized quasiprobabilities, i.e., Gaussian filtered $P$~functions, smoother than the Wigner function to visualize the quantum effects in the presence of high losses. 

\begin{figure}[t]
	\centering
	\includegraphics[width=0.8\linewidth]{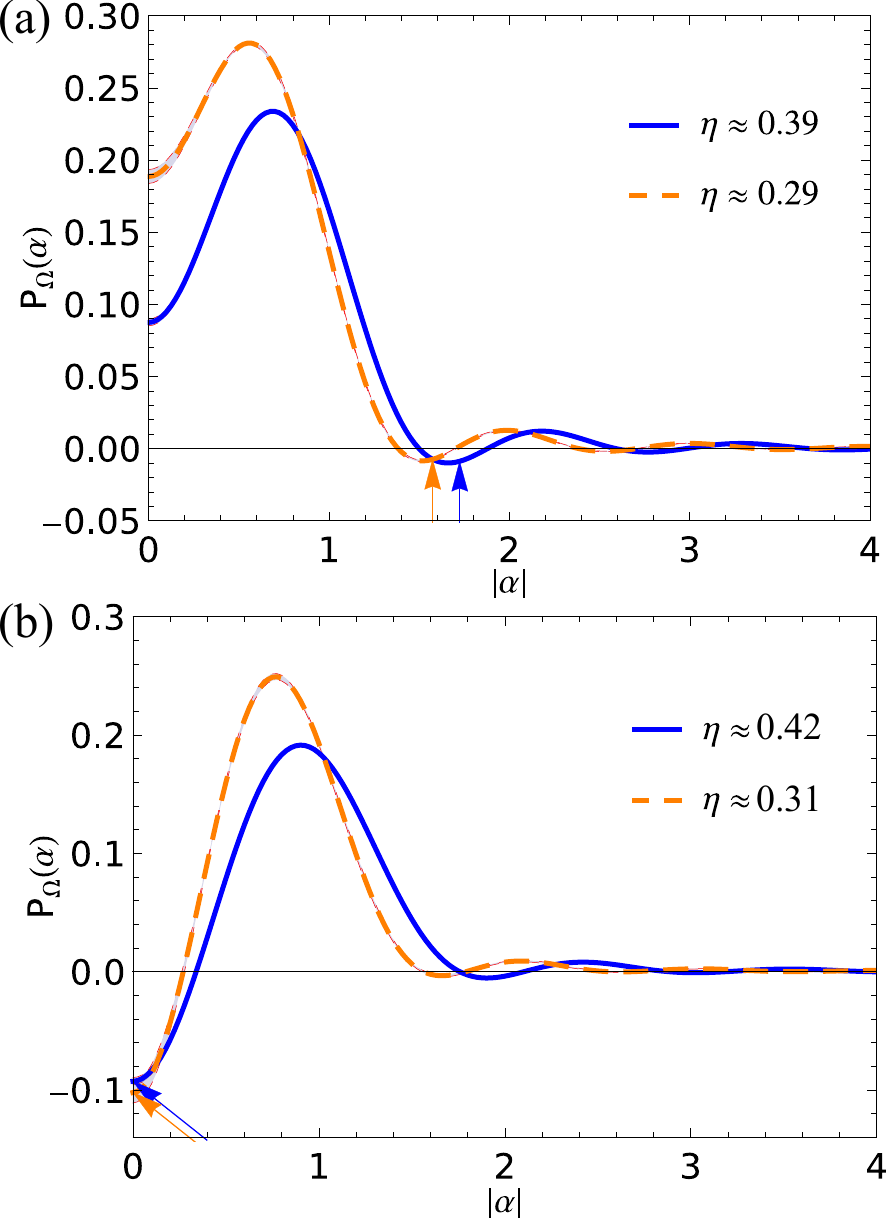}	
		\caption{
			(Color online)
			Nonclassicality quasiprobability as a function of $|\alpha|$, for the filter parameters $w_{\mathrm{opt}}$ which maximizes the statistical significance of the negativities; see Tab.~I in~\cite{supp}.
			They are shown for the two largest quantum efficiencies.
			The $|\alpha|$-value for which the negativity with the maximal significance is obtained is marked by an arrow, and its position depends on $w_{\mathrm{opt}}$, which is independently optimized for each state.
			(a) Single photon; (b) Two photons.
			The thin dashed line corresponds to an error of one standard deviation, which is barely noticeable.
		}
		\label{fig:NQ}
	\end{figure}

 	For comparison we utilize the non-Gaussian filter in Eq.~\eqref{eq:FilterInfty} for various values of the filter parameter $w$ into our direct sampling formula [Eqs.~\eqref{eq:directinsensitive} and~\eqref{eq:phaseindependentsigma}] to get the nonclassicality quasiprobabilities $P_\Omega(\alpha)$. 
	In Figs.~\ref{fig:NQ}~(a) and (b) the nonclassicality quasiprobabilities corresponding to the optimal filter parameters are shown for the efficiencies considered, see Supplemental Materials~\cite{supp} for more details. 
	We certify nonclassicality for the efficiencies of about $0.4$ and $0.3$ for both the single-photon state and the two-photon state with a very high statistical significance of more than 11 standard deviations, see~\cite{supp}.
	The results show that for the same number of data points a smaller quantum efficiency leads to a smaller maximal significance which is obtained for a larger optimal filter parameter.
	Since the statistical significance increases as the square root of the number of data points, by increasing the latter, i.e., enlarging the measurement time, it is possible for all efficiencies under study to arbitrarily reduce the statistical error of the negativities of the nonclassicality quasiprobabilities and, thus, to significantly certify nonclassicality.
	
\paragraph*{Conclusions.---}\hspace{-3ex}
In conclusion, we demonstrated that nonclassicality quasiprobabilities clearly outperform $s$-parametrized quasiprobabilities in uncovering signatures of nonclassicality of light in phase space in the presence of high losses. 
The regular $s$-parametrized quasiprobabilities, being Gaussian filtered versions of the Glauber-Sudarshan $P$~function, completely fail to indicate nonclassicality by negative values in many cases, such as for quantum efficiencies smaller than $0.5$.
By contrast, the nonclassicality quasiprobabilities, which are always regular functions obtained by proper non-Gaussian filtering of the $P$-function, enable a \textit{universal} nonclassicality test.
Our results are based on a real experiment where we generated single and two photon states with high losses and detected the produced light with phase-insensitive balanced homodyne detection. 
Specifically adapted to this measurement scenario we developed a direct sampling formula to retrieve the regular quasiprobabilities from a set of quadrature samples.
The results of this work underline the supremacy of nonclassicality quasiprobabilities to regularize the $P$~function and to certify all possible nonclassical effects of light. Note that it has been shown most recently that combining different $s$-parameterized quasiprobabilities can improve their potential to verify nonclassicality~\cite{Bohmann2020,biagi2020}.
		

\paragraph*{Acknowledgements.---}\hspace{-3ex}
This project has received funding from the European Union's Horizon 2020 research and innovation programme under Grant Agreement No. 665148, and the National Science Foundation under Grant No. 1620822. The authors would like to thank Jan Sperling for valuable discussions.
\begin{acknowledgements}
\end{acknowledgements}
	
\pagebreak

\section{Supplemental Materials}

\subsection{Experiment}
\label{sec:experiment}
In this section we use our methods to characterize light prepared in single- and two-photon states.
The general experimental scheme is shown in Fig.~1 of the main text.
It is based on Ref.~\cite{Cooper2013_2} with slight improvements. 
Light from a commercial Titanium-Sapphire femtosecond oscillator (Tsunami, Spectraphysics) is separated by a highly-transmissive beam-splitter into the local oscillator (LO) and a strong beam utilized to generate the quantum light. 
The latter is frequency-doubled in a Bismuth Borate (BBO) crystal with a thickness of 0.7 mm. 
The second harmonic is spatially-filtered to create a clean Gaussian beam with an efficiency of 0.25. 
The remaining 120 mW of the beam are used to pump a parametric down-conversion process (PDC) in a bulk potassium dihydrogen phosphate (KDP) crystal of 8 mm length. 
This source creates orthogonally-polarized photon pairs which are non-entangled in frequency~\cite{Mosley2008}, hence the purity of the state is close to unity. 
The daughter photons are separated into a signal and idler path on a polarizing beamsplitter. 
The idler, which has a spectral bandwidth of 12 nm full-width half maximum (FWHM), is used to herald the signal (2.5 nm FWHM). 
The heralding is achieved using a pseudo-number resolving photon detector with a fiber beamsplitter (FBS) and two superconducting avalanche photodiodes (Perkin-Elmer). 
Eventually, the signal field is combined with the LO on a 50-50 beamsplitter and detected with a fast balanced homodyne detector~\cite{Cooper2013} allowing to resolve individual pulses.

The LO is spatially filtered with a short single-mode fiber (SMF) and then spectrally filtered with an interference filter to match the spectral bandwidth of the signal photon. 
Also, a retro-reflector mounted on a linear translation stage is used for coarse matching of the delay between the two homodyne fields, and a mirror-mounted piezo-electric actuator is utilized to sweep the phase. 
A standard acquisition consists of letting the relative phase drift and trigger the detection on either one or two photon events with the heralding and accumulating the time traces, as shown by the inset of Fig.~1 in the main text. 


In the present work, we aim to study the phase-insensitive direct sampling of regular phase-space representations of the state for different detection efficiencies. 
This global detection efficiency may be written as the product,
\begin{align}\label{eq:etatot}
\eta = \eta_\textrm{bhd} \cdot \eta_\textrm{mm} \cdot \eta_\textrm{p} \cdot \eta_\textrm{dn},
\end{align}
of multiple contributions.
Here $\eta_\textrm{bhd}$ is the efficiency of the balanced homodyne detector, which includes losses after the 50-50 beamsplitter and the quantum efficiency of the detector.
The efficiency $\eta_\textrm{p} = \sqrt{\mathcal{P}}$ is related to the spectral purity $\mathcal{P}$ of the heralded state and $\eta_\textrm{dn}$ incorporates the dark counts of the detection that can lead to false positive events.
Imperfect overlap between the LO and the signal field is taken into account by the efficiency $\eta_\textrm{mm}$. 
Every quantities were evaluated precisely in Refs.~\cite{Cooper2013_2} and~\cite{Cooper2013}. 
We found that $\eta_\textrm{bhd} = 0.86$ and $\eta_\textrm{dn} \approx 1$ because the dark counts are negligible compared to the real events. 
Concerning the efficiency $\eta_\textrm{p}$, while the polarization and spatial purity remains unchanged from Ref.~\cite{Cooper2013_2} thanks to the filtering of the herald, additional work allowed for an improved measurement of the spectro-temporal purity. 
Notably, we utilized the measurement of the full joint spectral amplitude of a similar source as presented in Ref.~\cite{Davis2018} to evaluate that purity which encompasses correlations in the spectral phase. 
We measured a purity of $\mathcal{P} \approx 0.87$, yielding a lower value of $\eta_\textrm{p} \approx 0.94$. 

Finally, the measurement of the overlap $\eta_\textrm{mm}$ between the two modes was slightly improved as well. 
Seeding the source with a bright coherent state results in difference frequency generation (DFG) of an orthogonally-polarized beam whose profile should resemble that of the squeezed vacuum. 
Using the DFG beam, it is possible to optimise the contrast of the interference fringes with the LO. 
Using a free-space detector and equal power in the DFG and LO, we measured a contrast of 86\%, which represents exactly the product of the overlap of every degree of freedom of the field, i.e., polarisation, spatial and temporal. 
The spatial contribution can be removed by monitoring the contrast after filtering by a SMF, yielding a contrast of 90\%. 
Therefore, the spectral-temporal overlap is estimated at $0.96$. 
Lastly, the heralding efficiency is required to deduce the final modematching efficiency. 
By monitoring single photon coincidences between the signal and the idler, we obtained a maximum heralding efficiency of $0.23$, which is $0.51$ when compensating for the quantum efficiency $0.45$ of the single photon detectors. 
This measurement with single photons allows us to take into account the difference in spatial profile between the DFG and the squeezed vacuum beam. 
Computing the product between the heralding efficiency and the spectral-temporal overlap, we have $\eta_\textrm{mm} \approx 0.49$. 
Accordingly, the global efficiency of the detection [Eq.~\eqref{eq:etatot}] is then estimated to be $\eta \approx 0.4$.
This number is lower than that presented in Ref.~\cite{Cooper2013_2}, which was $0.54$, because of the reduced heralding efficiency. 
This is likely due to the fact that the source was aligned to deliver many more single photon events: $500$ kHz in our case compared to $180$ kHz in Ref.~\cite{Cooper2013_2}.

\begin{figure}[t]
	\centering
	\includegraphics[width=.86\linewidth]{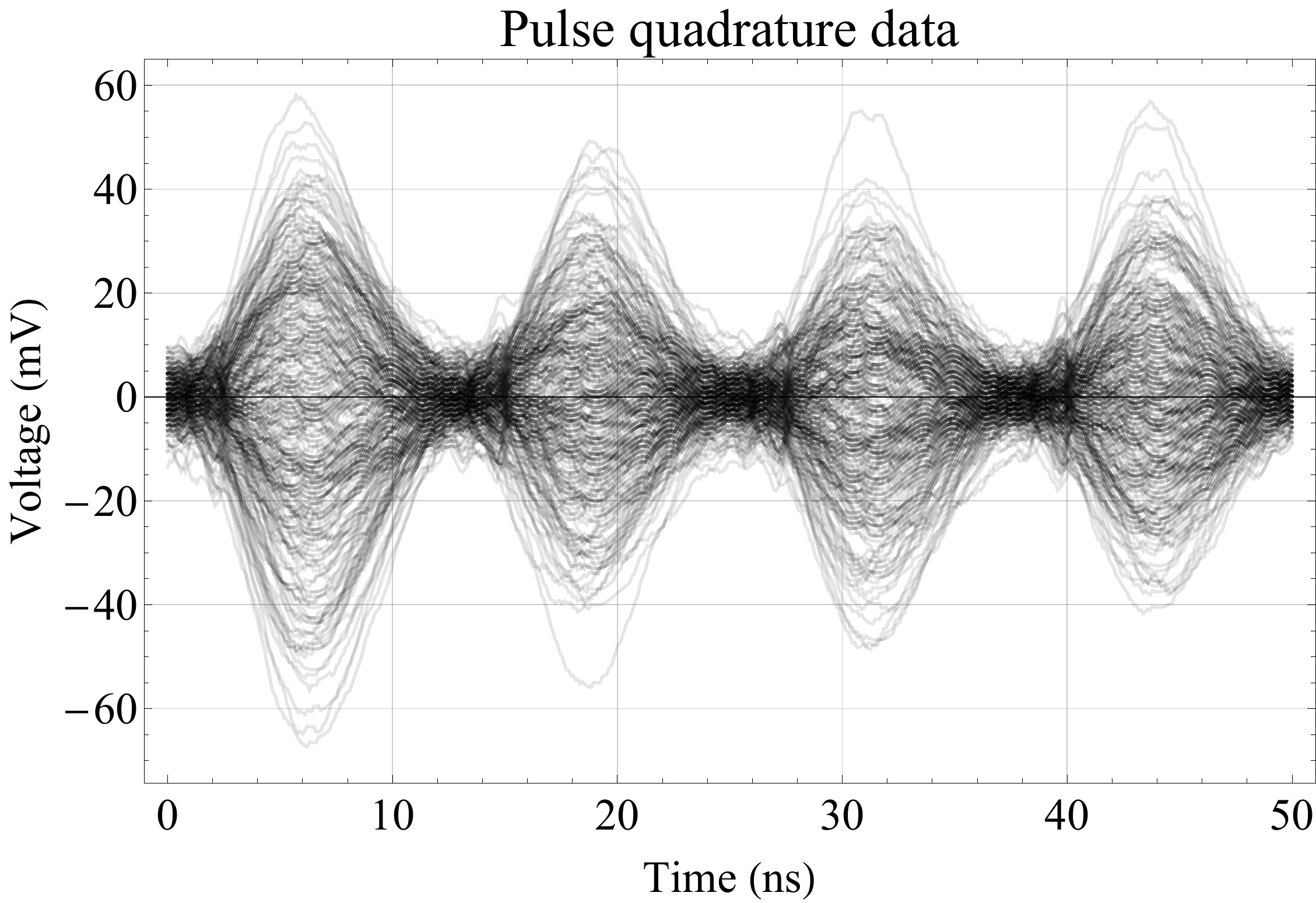}
	\caption{BHD voltage corrected for residual signal for 250 acquisitions.}
	\label{fig:pulses}
\end{figure}

A single quadrature point consists of computing the statistics of a certain number of acquisitions (inset of Fig.1 in the main text) from which the residual mean voltage can be safely removed to obtain the pulses shown in Fig.~\ref{fig:pulses}. 
The triggering is set such that the first pulse of the acquisition contains the photon quadrature data, whereas the next three pulses are mostly in a vacuum state. 
Computing the statistics of each pulse over a certain number of acquisitions yields one quadrature data point for the Fock state and three redundant data points for vacuum.
These are obtained by integrating each pulse photocurrent over their FWHM to avoid errors due to sampling. 
The vacuum data is used to normalize the quadrature variance of the vacuum state to one.
We sampled $8\cdot10^5$ and $6\cdot10^5$ quadrature points for single and two photon states, respectively. 
The trigger rate was 500 kHz in the former case and 500 Hz in the latter. 

\begin{figure}[t]
	\centering
	\includegraphics[width=0.95\linewidth]{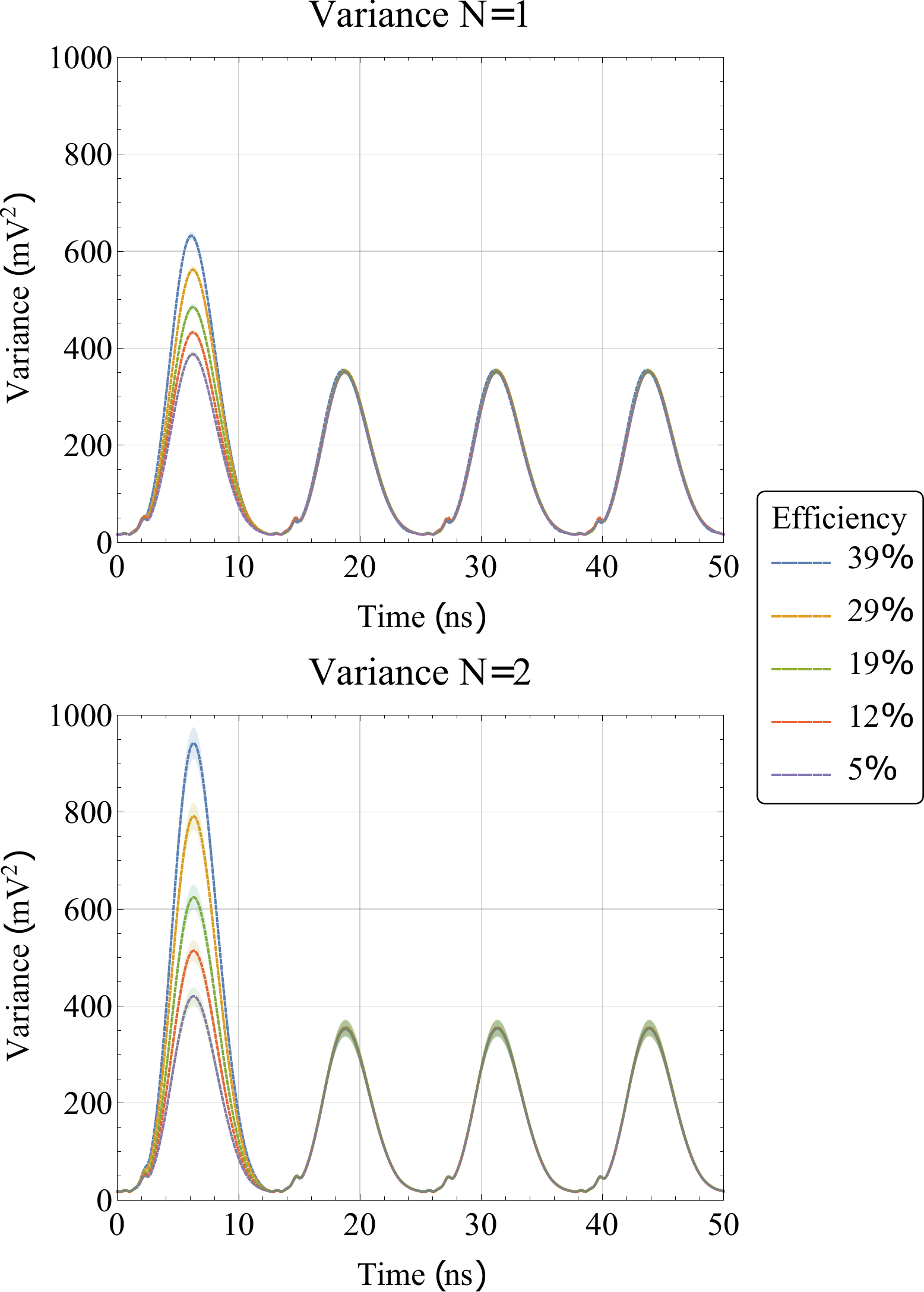}
	\caption{(Color online) Variance of the photocurrents from Fig.~\ref{fig:pulses} over the full data set for single (top) and two photons (bottom5) states. Shadowed trace represent the standard deviation over the dataset.}
	\label{fig:var}
\end{figure}

A visual representation of the measurement may be obtained by computing the variance of the photocurrent over a number of acquisition, which is depicted in Fig.~\ref{fig:var} for the full data set of single and two photon quadrature data. 
We can see that the first pulse indeed contains a higher photon number than the three next pulse which have an equal variance. 
Also, the variance for a two photon state is apparently larger than that for a single photon state. 
The magnitude for the first pulse also decreases with the efficiency, as expected. 

If the Fock state $|n\rangle$ undergoes constant loss, described by the quantum efficiency $\eta$ the resulting state
\begin{align}\label{eq:pneta}
\hat\rho(\eta;n)=\sum_{k=0}^n\binom{n}{k}\eta^k(1-\eta)^{n-k}|k\rangle\langle k|
\end{align}
is a mixture of Fock state with photon number smaller than or equal to $n$.
This resulting state is phase-independent and its normalized quadrature variance reads as
\begin{align}\label{eq:varx}
\mathrm{var}\,\hat x=2n\eta+1.
\end{align}
Accordingly, one way to estimate the quantum efficiency without performing a full state reconstruction is by using the formula
\begin{align}\label{eq:etaest}
\eta = \dfrac1{2n}\left(\mathrm{var}\,\hat x-1\right).
\end{align}
Since the single photon data is acquired at a high trigger rate, that efficiency can be monitored in real time. 
This allows for a fine tuning of the overlap between the signal and the LO using squeezed vacuum rather than the DFG beam. 
A similar development for the two photon case also allows to extract the efficiency. 
For the full data set, we showed that the efficiency remained constant during the acquisition thanks to careful design of the experiment. 
\begin{figure}[b]
	\centering
	\begin{minipage}[h]{0.80\linewidth}
		(a) Single photon\\
		\includegraphics[width=0.99\linewidth]{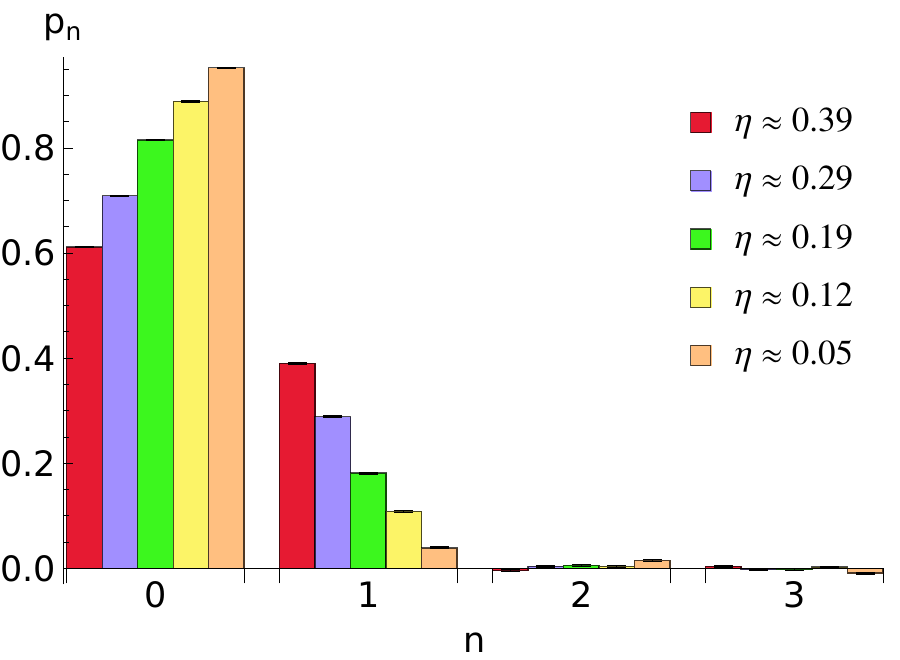}	
		\\\vspace*{1ex}(b) Two photons\\
		\includegraphics[width=0.99\linewidth]{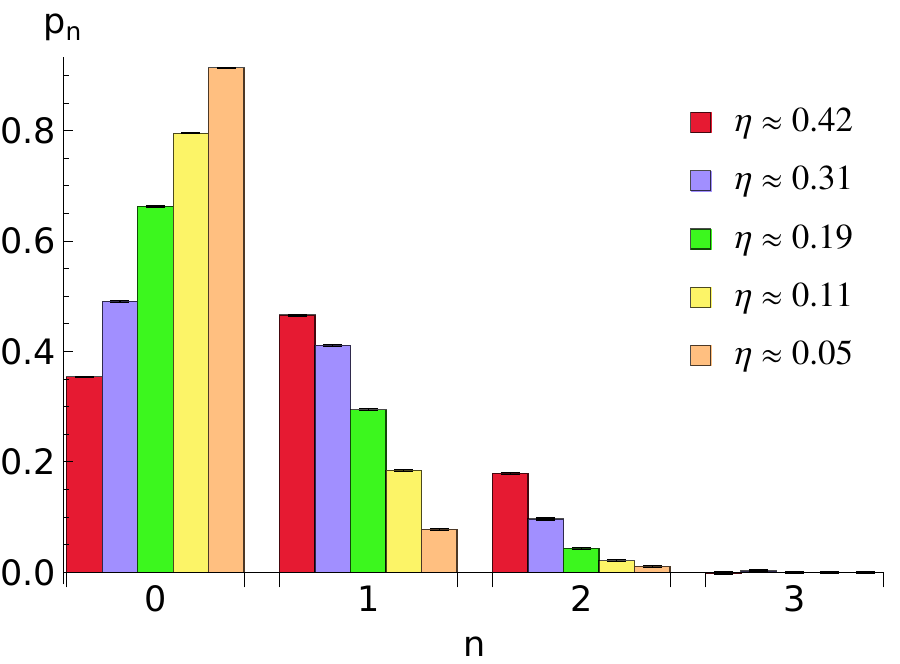}
	\end{minipage}
	\caption{
		(Color online)
		The sampled photon-number distribution for (a): Single photon state; (b): Two photon state measured with quantum efficiency $\eta$. 
		The error bars indicate an uncertainty of one standard deviation.			
	}
	\label{fig:pnd}
\end{figure}

\begin{table}[t]
	\centering
	\label{tab:opt}
	\begin{tabular}{c|ccccc|ccccc}
		\hline\hline
		&\multicolumn{5}{c}{Single photon}&\multicolumn{5}{c}{Two photons}\\		
		\hline
		$\eta$&$0.39$&$0.29$&$0.19$&$0.12$&$0.05$&$0.42$&$0.31$&$0.19$&$0.11$&$0.05$\\
		$N (\times 10^5)$&$8$&$8$&$8$&$8$&$8$&$6$&$6$&$6$&$6$&$6$\\
		$\left|\Sigma_{\mathrm{max}}\right|$&$51.8$&$23.2$&$3.7$&$2.1$&$4.9$&$29.6$&$11.5$&$1.6$&$1.1$&$0.7$\\
		$w_{\mathrm{opt}}$&$1.50$&$1.65$&$1.90$&$2.35$&$2.20$&$1.55$&$1.75$&$2.50$&$2.50$&$2.50$\\
		\hline\hline
	\end{tabular}
	\caption{Results for the single- and two-photon measurements each performed for five different quantum efficiencies $\eta$ listed in five columns.
		Each column contains the estimated quantum efficiency $\eta$, the number of recorded data points $N$, and the maximal statistical significance $|\Sigma_{\mathrm{max}}|$ of the negativities of the directly sampled nonclassicality quasiprobability obtained for the optimal filter parameter $w_{\mathrm{opt}}$.
	}
\end{table}

The relative delay between the signal and the LO was stepwisely changed to reduce the efficiency in order to study the performance of our sampling formulas of regular phase-space distributions for high losses. 
For a signal in the spectral-temporal mode $u_s$ and the LO in the mode $u_\textrm{LO}$, the spectral overlap between these two modes is simply given by $\eta_\mathrm{spec} = \mathrm{Re}\{\int \mathrm{d}\omega \, u_s^\ast(\omega) u_\mathrm{LO}(\omega)\}$, which was estimated above as $0.96$ at zero delay. 
Hence, introducing a larger delay will decrease the overlap following the temporal envelope of the cross-correlation between the signal and LO modes. 
This has the advantage of being easily adjustable and reversible.
As a result, we acquired quadrature data for single- and two-photon states for five different efficiencies, estimated by Eq.~\eqref{eq:etaest}. 
They are listed in Tab.~\ref{tab:opt} and are all smaller than 0.5. 

\begin{figure}[t]
	\centering
	\begin{minipage}[t]{0.80\linewidth}
		(a)\\
		\includegraphics[width=0.99\linewidth]{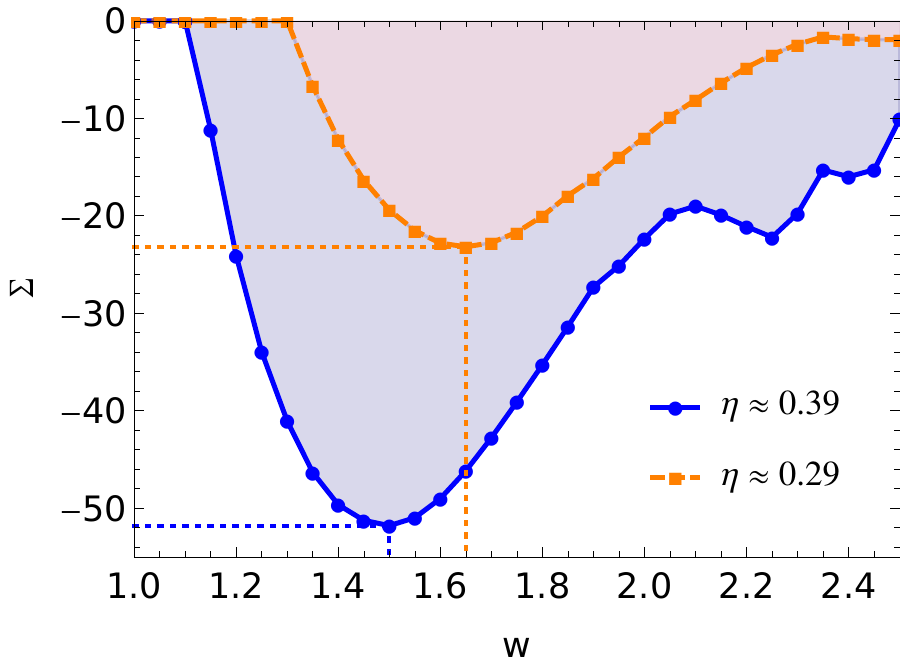}	
		\\\vspace*{1ex}(b)\\
		\includegraphics[width=0.99\linewidth]{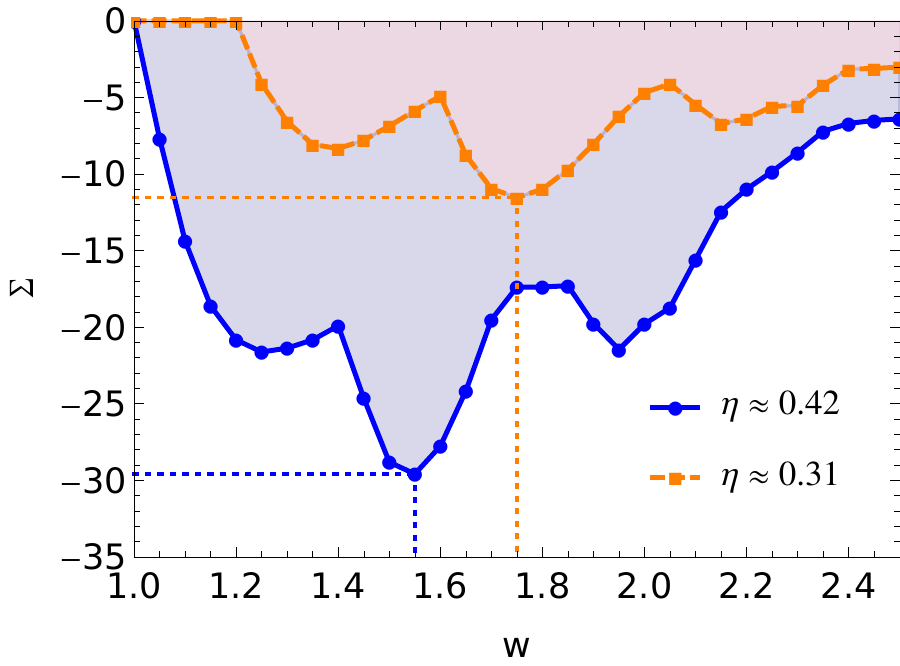}
	\end{minipage}
	\caption{
		(Color online)
		Signed significance $\Sigma$ of the negativity of the sampled nonclassicality quasiprobabilities as a function of the filter parameter $w$ for the two largest quantum efficiencies.
		The markers correspond to the actually tested parameters of $w$.
		(a): Single photon state; (b): Two photon state. 				
		The dashed lines mark the maximal significance obtained for the optimal filter parameter $w_{\mathrm{opt}}$; see also Tab.~\ref{tab:opt}.
	}
	\label{fig:sigplot}
\end{figure}

To further characterize the probed light, we additionally sampled the photon-number distribution $p_n=\langle n|\rho|n\rangle$ from the quadrature data with the techniques which have originally been applied to experiments in Ref.~\cite{Munroe95}. This yields the results in Fig.~\ref{fig:pnd}~(a) and ~(b) for the single photon and the two photon case.	
As expected, photon numbers larger than the heralded number of photons barely contribute.
Furthermore, the statistics roughly reflects the expected distribution of a Fock state measured with quantum efficiency $\eta$ [cf. Eq.~\eqref{eq:pneta}].
Possible discrepancies are due to the heralding detector, which is not perfectly photon-number resolving.

\subsection{Statistical significance of nonclassicality}
\label{sec:significance}
In the present section we provide the details on the experimentally determined statististical significances of the verification of the nonclassicality by the non-Gaussian filtered quasiprobabilities. We determined the signed statistical significance $\Sigma$. 
The results for the two largest quantum efficiencies for the single- and two-photon state are depicted in Fig.~\ref{fig:sigplot} (a) and~(b).
A negative value of $\Sigma <-5$ corresponds to a significant verification of the negativity of $P_\Omega$, and hence, of the nonclassicality of the quantum state under study.
In statistics it is often useful to refer to a value of $|\Sigma|$ larger than $5$ to be statistically significant. This is clearly the case for the states prepared in our experiments shown in Fig.~\ref{fig:sigplot}.
For an optimal filter width, $w_{\mathrm{opt}}$, we obtain a maximal statistical significance $|\Sigma_{\mathrm{max}}|$.
These values are listed in Tab.~\ref{tab:opt}. Note that while the single-photon measurement with $\eta\approx 0.05$ yields a quite large significance, the photon-number distribution of this state does not perfectly follow the distribution of a lossy single photon, but also contains contributions of a two-photon state (see Fig.~\ref{fig:pnd}).

\bibliography{biblio}

\end{document}